\documentclass[prd,preprint,eqsecnum,nofootinbib,amsmath,amssymb,
               tightenlines,floatfix]{revtex4}
\input epsf
\usepackage{color}
\usepackage{bm}
\def\putbox#1#2{\epsfxsize=#1\textwidth\epsfbox{#2}}
\def\centerbox#1#2{\centerline{\epsfxsize=#1\textwidth\epsfbox{#2}}}
\def\Tr{\,{\rm Tr}\:}

\def\be{\begin{equation}}
\def\ee{\end{equation}}
\def\st{\be}
\def\stp{\ee}
\def\bea{\begin{eqnarray}}
\def\eea{\end{eqnarray}}
\def\sss{\scriptscriptstyle \rm }

\def\FFdual{\frac{g^2}{32\pi^2}F^a_{\mu\nu} \tilde{F}^{\mu\nu}_b}
\def\FFddual{\frac{g^2}{32\pi^2}F^a_{\alpha\beta} \tilde{F}^{\alpha\beta}_b}

\def\Im{\,{\rm Im}\:}
\def\Eq#1{Eq.~(\ref{#1})}

\def\NCS{N_{_{\rm CS}}}
\def\ncs{\NCS}

\def\OO{{\cal O}}
\def\taul{{\tau_{\sss L}}}
\def\tham{{t_{\sss H}}}
\def\thamsq{{t_{\sss H}^2}}

\def\x{{\bm x}}
\def\y{{\bm y}}
\def\p{{\bm p}}

\def\FF{\tilde F}

\def\nf{N_{\rm f}\,}
\def\nc{N_{\rm c}}
\def\mD{m_{\sss D}}

\def\alphas{\alpha_{\rm s}}

\def\lsim{\mbox{~{\raisebox{0.4ex}{$<$}}\hspace{-1.1em}
        {\raisebox{-0.6ex}{$\sim$}}~}}

\advance\parskip 2pt

\begin{document}

\title
    {
      The Sphaleron Rate in SU($N$) Gauge Theory
    }

\author{Guy D.~Moore and Marcus Tassler}
\affiliation
    {%
    Department of Physics,
    McGill University,
    3600 rue University,
    Montr\'{e}al, QC H3A 2T8, Canada
    }%

\date {November 2010}

\begin {abstract}%
    {%
The sphaleron rate is defined as the diffusion constant for topological
number $\NCS \equiv \int \frac{g^2F\FF}{32\pi^2}$.  It establishes the
rate of equilibration of axial light quark number in QCD and is of
interest both in electroweak baryogenesis and possibly in heavy ion
collisions.  We calculate the weak-coupling behavior of the SU(3)
sphaleron rate, as well as making the most sensible extrapolation
towards intermediate coupling which we can.  We also study the behavior
of the sphaleron rate at weak coupling at large $\nc$.
    }%
\end {abstract}

\maketitle
\thispagestyle{empty}

\section{Introduction}

It has long been known \cite{Polyakov} that Yang-Mills theory
possesses multiple classical vacua, distinguished by a topological
number, the Chern-Simons number.  No continuous sequence of
infinitesimal gauge transformations can bring one of these vacua into
another; instead a ``large'' gauge transformation of nontrivial topology
is needed.  The topological character of the gauge transformations is
ensured by the nontrivial homotopy of the gauge group, {\it e.g.\ }
$\pi_3(G) = {\cal Z}$ for $G={\rm SU}(\nc)$, $\nc \geq 2$.

This topological nature of the vacuum has important physical
consequences because of the chiral (ABJ) anomaly
\cite{Adler,BellJackiw}.  The would-be conserved current associated with
a left-handed fermion in the fundamental representation is%
\footnote{%
    Our conventions are that
    $g_{\mu\nu} = \eta_{\mu\nu} = {\rm Diag}[\,{-}\,{+}\,{+}\,{+}]$,
    and that capital letters $X,P$ are 4-vectors with
    $t=x^0$, $\omega=p^0$ and $\x,\p$ are the 3-vector components.
    Lower case $x,p$ are reserved to denote $|\x|,|\p|$.
 }
\st
\label{current1}
\partial_\mu J^\mu_{\sss L}(X) = \FFdual(X) \,, \qquad
\FF^{\mu\nu} \equiv \frac 12 \epsilon^{\mu\nu\alpha\beta}
  F_{\alpha\beta}\,.
\stp
Within the Standard Model, this has a few very interesting
consequences.  The ${\rm SU}_{\sss L}(2)$ weak-hypercharge group couples
only to left-handed fields, so the anomaly gives rise to baryon and
lepton number violation \cite{tHooft}.  While under ordinary conditions
these have essentially no physical consequences, at very high
temperatures where electroweak symmetry is ``restored,'' baryon and
lepton number violation becomes efficient
\cite{Manton,Rubakov,ArnoldMcLerran}, opening the possibility of
electroweak baryogenesis \cite{baryoreview}.
The ${\rm SU}_{\sss C}(3)$ color group couples to equal numbers of left
and right handed fields%
\footnote{or equivalently, to equal numbers of left-handed fields in the
  fundamental and antifundamental representations}
and so no global symmetry violation is associated with it.  However, the
up and down quark flavors are extremely light, so the anomaly is the
most efficient way of interchanging left with right handed first
generation quarks.

The equilibration of right- and left-handed quark number is important to
baryogenesis, and therefore the strong sphaleron rate is too;
see for instance \cite{Giudice,ewbg1,ewbg2,ewbg3}.  The same
applies to heavy ion collisions, which also appear to involve the
production of a hot quark-gluon plasma \cite{qgp_obs}.  Indeed,
recently Kharzeev, McLerran and Warringa have argued that topological
fluctuations may play an important role in {\sl creating} local
fluctuations or imbalances in left-handed versus right-handed quark
number \cite{Kharzeev1,Kharzeev2}, which could have very interesting
experimental signatures \cite{Voloshin}.

The relation between real (Minkowski) time topology change and Euclidean
topological susceptibility is subtle
(see \cite{strikesback} and the discussion below).  Therefore the
Minkowski rate of topology fluctuation cannot be easily inferred from
Euclidean quantities and needs to be separately studied.  There is a
vast literature on doing so for the ${\rm SU}_{\sss L}(2)$ sector of the
Standard Model \cite{Rubakov,ArnoldMcLerran,Ambjorn,BodekerMcLerran,
AmbjornKrasnitz,MooreTurok,HuMuller,ASY1,Bodeker,ASY2,BMR,MooreB,
ArnoldYaffe}, and it would be fair to say that the efficiency
of topology change is well understood for this group and at the relevant
coupling.  But ${\rm SU}_{\sss C}(3)$ has been much less carefully
studied.  At this time, there are two quite different parametric
estimates \cite{Giudice,MooreSEWM2K} and one rather crude
calculation \cite{MooreSU3} which was made before our modern
understanding of the parametric behavior was developed
\cite{Bodeker,ASY2}.  Therefore, particularly in the light of the recent
resurgence of interest in topology change in the strong interactions
\cite{Kharzeev1,Kharzeev2}, we think it is timely to revisit the
question of the ``strong sphaleron rate,'' the (equilibrium, thermal)
rate of topology change in the group SU(3).  As we will discuss
momentarily, we currently only know of tools to do so at weak coupling,
so we will be forced to restrict our attention to weak coupling.
However we will also make the most reasonable estimate we can of the
rate at intermediate coupling.

For the impatient reader, we summarize our results here:
the rate of topology change per spacetime volume is defined as
\be
\Gamma_{\rm sphal} \equiv \lim_{t\rightarrow \infty}
\frac{(\ncs(t) - \ncs(0))^2}{Vt}
= \int d^4 X \left\langle \FFdual(X) \; \FFddual(0)
\right\rangle
\ee
where the $t\rightarrow \infty$ (zero frequency) limit is to be taken
with some care, see below.  In the limit of high temperature (and hence
of weak coupling), the temperature and coupling dependence for SU(3) is
($\alphas = g^2/4\pi$)
\bea
\label{mainresult}
\Gamma_{\rm sphal} &=& (132 \pm 4) \left( \frac{g^2 T^2}{\mD^2} \right)
\left( \ln \frac{\mD}{\gamma} + 3.0410 \right) \alphas^5 T^4 \,,
\\
\gamma & = & \frac{\nc g^2 T}{4\pi} \left( \ln \frac{\mD}{\gamma}
+ 3.041 \right) \,,
\\
\mD^2 & = & \frac{2\nc + \nf}{6} g^2 T^2 = \frac{6+\nf}{6} g^2 T^2 \,.
\eea
Note that $\gamma$, the mean rate of color randomization, is defined
self-referentially, and should be found self-consistently
\cite{ArnoldYaffe}.

The $\nc$ scaling is surprisingly simple; investigating
$\nc=2,3,5,8$ we find that
\bea
\label{ncscaling}
\Gamma_{\rm sphal} & = & (0.21 \pm 0.01)
\left( \frac{\nc g^2 T}{\mD^2} \right)
\left( \ln \frac{\mD}{\gamma} + 3.0410 \right)
\frac{\nc^2 - 1}{\nc^2}
(\nc \alpha)^5 T^4
\eea
with no further $\nc$ corrections visible within our error bars.  The
functional form and even the numerical value are in
good agreement with the conjecture of \cite{MooreSEWM2K}.

The formal range of validity of the above results is very narrow; not
only the coupling $\alphas$, but even the inverse log of the coupling
$1/\ln(1/\alphas)$ must be small.  At larger coupling we can {\sl
  estimate} the sphaleron rate based on the diffusion rate of a
lattice-regulated, classical field theory with the same magnetic-field
damping rate as the leading-order result we find in the quantum theory.
This approach is known to be correct in the weak coupling limit
\cite{Bodeker,Arnoldlatt,ASY2} but at larger coupling values it amounts
to a crude estimate; the dynamics cease to be well described by
classical field theory, and the classical lattice model sees more and
more ``lattice-y'' dynamics artifacts.  We can also only push the
lattice to a certain level of coarseness before it becomes impossible to
establish any topological character for the fields.  As shown in
Table \ref{table_finitea}, this allows access to $\alphas \lsim 0.1$.
The leading-log results work reasonably well for $\alphas \lsim 0.03$
but break down after this.  And the classical framework starts to break
down above about $\alphas = 0.1$.

The remainder of the paper is organized as follows.  In Section
\ref{sec:generalities} we review the definition and utility of the
sphaleron rate, its relation to Euclidean correlation functions, and the
extreme difficulty of its determination from lattice Euclidean data.
Then Section \ref{sec:bodeker} reviews the effective field theories
valid at weak coupling and presents the details of our numerical
implementation and calculation. Section \ref{results} presents our
numerical results for the weak-coupling sphaleron rate and Section
\ref{finite_a} presents the best extrapolation towards intermediate coupling
we can achieve.  We discuss applications in Section \ref{discussion}.

\section{Sphaleron rate:  generalities}
\label{sec:generalities}

Here we review the definition and physical relevance of the sphaleron
rate, and its relation to Euclidean correlation functions.  Nothing in
this section is new, it is included as a refresher for readers
familiar with the issues and an introduction for readers who are not.

\subsection{Definition of the sphaleron rate}

As mentioned above, the axial current associated with a quark species,
\st
J^\mu_{A,q} \equiv \bar q \gamma^\mu \gamma_5 q
\stp
is not conserved.  There are actually two contributions to its
nonconservation:  the quark mass, which explicitly breaks the axial
symmetry, and the chiral anomaly:%
\footnote{From now on we specialize to vectorlike theories with
  fundamental representation matter.}
\st
\label{current2}
\partial_\mu J^\mu_{A,q} = 2 m_q \bar q \gamma_5 q
- 2\FFdual
\stp
(where the factor of $-2$ difference from \Eq{current1} is because
$J^\mu_A$ is the right-handed minus the left-handed currents).
In real-world QCD the up and down quark masses are very small,
$m_{u,d} \simeq 2.0,4.5\:{\rm MeV}$ \cite{Bazavov:2009tw} and so in many
situations one may neglect the explicit current nonconservation.

Neglecting the quark masses, the amplitude to evolve from state
$| \psi_1(t_1)\rangle$ to state $| \psi_2(t_2)\rangle$ {\sl times} the
change in axial number is (in the Heisenberg picture)
\st
A\Delta Q_{A,q} = 2 \int_{t_1}^{t_2} dt \int d^3 \x
\langle \psi_2 | \FFdual(\x,t) | \psi_1 \rangle \,.
\stp
The probability of the process times the square of the change is
\st
P (\Delta Q_{A,q})^2 = 4 \int d^4 X d^4 Y
\langle \psi_1 | \FFdual(X) |\psi_2 \rangle \langle \psi_2 |
\FFddual(Y) | \psi_1 \rangle \,.
\stp
Summing over final states,
$\sum_{\psi_2} |\psi_2 \rangle \langle \psi_2 | = 1$, we find
\st
\langle (\Delta Q_{A,q})^2 \rangle = 4
\int d^4 X d^4 Y \langle \psi_1 | \FFdual(X) \FFddual(Y)
|\psi_1 \rangle \,.
\label{meansqr}
\stp
The mean squared change for a general density matrix $\rho$ is found by
replacing $(\langle \psi_1| \ldots |\psi_1 \rangle)$ with
$({\rm Tr}\: \rho \ldots)$.  Note that the operators
here are not time-ordered;  it is the Wightman correlator
\st
G^>_{F\FF}(X,Y) \equiv \left\langle \FFdual(X) \; \FFddual(Y)
\right\rangle
\stp
which is relevant here.

\Eq{meansqr} holds for arbitrary density matrices including nonequilibrium
density matrices.
Here we will more modestly try to evaluate it for
the thermal ensemble.  Since the thermal ensemble is 4-translation
invariant, the mean-squared $Q_{A,q}$ change is extensive in 4-volume and it
is more useful to define a rate of change per 4-volume:
\bea
\langle (\Delta Q_{A,q})^2 \rangle_{\rm therm.} &=&
       4 Vt \, \Gamma_{\rm sphal} \,,
\nonumber \\
\Gamma_{\rm sphal} & \equiv & \int d^4 X G^>_{F\FF}(X,0) \,.
\label{def_sphal}
\eea
This is the same as the zero frequency and momentum limit of the
momentum-space Wightman function
\st
G^>_{F\FF}(P) \equiv \int d^4 X e^{-iP\cdot X}
G^>_{F\FF}(X,0) \,.
\stp

In defining the sphaleron rate in this way we have tacitly assumed that
the generated $(Q_{A,q})$ value can persist without influencing the
subsequent evolution of the system.  Of course this is not true; a net
abundance of $Q_{A,q}$ leads to chemical potentials for left and right
handed quark number which will then energetically bias future topology
change to allow that quark number to relax.  Defining
$Q_5 = \sum_q Q_{Aq}$ the total axial light quark charge, standard
fluctuation-dissipation arguments
\cite{KhlebnikovShaposh,Giudice} show that
\be
\frac{dQ_5}{dt} = - Q_5 \frac{(2\nf)^2}{\chi_Q}
\frac{\Gamma_{\rm sphal}}{2T}
\simeq -Q_5 \frac{6\nf}{\nc} \frac{\Gamma_{\rm sphal}}{T^3} \,,
\ee
where $\chi_Q$ is the susceptibility for $Q_5$ and in the last line we
used the leading-order perturbative estimate $\chi_Q = \nf \nc T^2/3$.
What this expression means is that any fluctuation-induced nonvanishing
value for $Q_5$ will bias all subsequent topology changing processes,
causing $Q_5$ to relax back to zero with fluctuations set by the
susceptibility.  This means that a fluctuation in topology of one sign
will eventually be balanced by a fluctuation of the other sign with a
time constant $\tau \sim 2T \chi_Q / (4 \nf^2 \Gamma_{\rm sphal})$.
To define the sphaleron rate we implicitly assume that this time scale
is long compared to any microscopic time scales, and we implicitly
cut off the time integration in \Eq{def_sphal} on a timescale shorter
than $\tau$ but longer than microscopic time scales.
At weak coupling the longest microscopic time scale is $\sim 1/\alphas^2
T$ while the relaxation time scale defined above is
$\sim 1/\alphas^5 T$ (as we shall see), so the separation of scales
exists.  The separation also exists at strong coupling in large $\nc$
${\cal N}{=}4$ SYM theory, where the sphaleron rate is
$\Gamma_{\rm sphal} = (g^2\nc)^2 T^4/256\pi^3 \sim \nc^0$
\cite{Son_SYM}, while $\chi_Q \sim \nc^2 T^2$.
Therefore $\tau \sim \nc^2 / T$ is
large in the large number of colors limit.%
\footnote{The calculation is only valid in the limit
  $\nc \gg (g^2 \nc) \gg 1$.  In this limit the relaxation time is
  parametrically large, while all microscopic timescales are $\sim T$.}
However for $\nc=3$ ordinary QCD at $T\sim 200$MeV it is by no means
clear that such a separation of scales exists, and it is possible that
one cannot really define a sphaleron rate {\it per se}.

\subsection{Relation to Euclidean correlation functions}

We saw above that the sphaleron rate is set by the zero frequency and
momentum limit of the $F\FF$ Wightman function.  Naively one might think
it is easy to determine the sphaleron rate from Euclidean correlation
functions, which can be measured on the lattice.  Indeed, in the late
1980's there were some misconceptions that such things should be
possible, which were quashed by a paper by Arnold and McLerran
\cite{strikesback}.  Here we remind the reader how the sphaleron rate is
related to Euclidean correlation functions, and why it would {\sl not}
be easy to determine it from Euclidean lattice data.  Though the remarks
here are rather generic to two-point functions and are well known,
we review them in a little detail because we are not
aware that they have been clearly discussed in this particular context.

The Wightman correlation function defined above is
\st
G^>_{F\FF}(X) \equiv \Tr e^{-\beta H}
   e^{iHt} \FFdual(\x) e^{-iHt} \FFddual(0)
\stp
while the Euclidean correlation function is
\st
G^{\sss E}_{F\FF}(\tau,\x) \equiv \Tr
e^{-\beta H} e^{\tau H} \FFdual(\x) e^{-\tau H} \FFddual(0) \,,
\stp
defined on the range $0<\tau<\beta$ and satisfying periodic boundary conditions,
$G^{\sss E}(\beta,\x) = G^{\sss E}(0,-\x) = G^{\sss E}(0,\x)$.%
\footnote{Note that $F\FF$ is a pseudoscalar. $G(-\x)$ can be related
          to $G(\x)$ by a rotation; parity is not involved.}
It is clear from the
definitions that $G^>(t,\x)$ is the analytic continuation
$\tau \rightarrow it$ of $G^{\sss E}(\tau,\x)$.  In particular the
equal-time value of $G^>$ equals the equal $\tau$ value of
$G^{\sss E}$.

\begin{figure}
\centerline{\begin{picture}(200,60)
  \thicklines
  \multiput(20,50)(40,0){4}{\vector(1,0){40}}
  \multiput(100,46)(40,0){2}{\color{blue}\vector(1,0){40}}
  \multiput(180,0)(-40,0){2}{\color{blue}\vector(-1,0){40}}
  \multiput(100,46)(0,-23){2}{\color{red}\vector(0,-1){23}}
  \put(20,40){$G^>(\omega)$}
  \put(140,6){{\color{blue}{$G^{\sss R}(\omega)$}}}
  \put(70,21){{\color{red}{$G^{\sss E}(\omega)$}}}
\end{picture}}
\caption{Integration contours in complex $t$ plane for
$G^>(\omega)$, $G^{\sss E}(\omega)$, and $G^{\sss R}(\omega)$
(the latter for ${\rm Im}(\omega)=2\pi n$ only).
Contour deformation turns $G^{\sss R}$, but not $G^>$, into $G^{\sss E}$.
\label{fig:cplane}}
\end{figure}
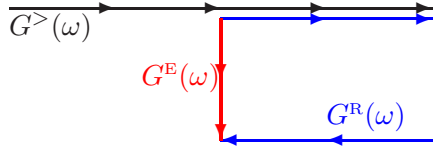

But what we want is the zero {\sl Minkowski frequency}
limit of $G^>$, and the frequency transforms of the two functions are
{\sl not} the same.  Instead, the frequency-space transform of the
Euclidean function continues to the retarded correlation function
\bea
G^{\sss R}_{F\FF}(t,\x) &\equiv& \left\langle \left[ \FFdual(t,\x)
\, , \, \FFddual(0,0) \right] \right\rangle \Theta(t) \nonumber \\
&=& \Theta(t) \left( G^>_{F\FF}(t,\x) - G^<_{F\FF}(t,\x) \right) \,.
\eea
The reason why is illustrated in Fig.~\ref{fig:cplane}:
first note that the frequency domain form of $G^{\sss R}$,
\st
G^{\sss R}(\omega) =
\int_0^\infty dt e^{i\omega t} \Big( G^>(t) - G^<(t) \Big)
\stp
is trivial to continue to complex $\omega$ with $\Im \omega>0$.  For
purely imaginary $\omega = i\omega_{\sss E}$ with
$\beta \omega_{\sss E} = 2\pi n$ it is
\st
G^{\sss R}(i\omega_{\sss E}) =
\int_0^\infty dt e^{i\omega_{\sss E} it} G^>(t)
- \int_0^\infty dt e^{i\omega_{\sss E} it} G^<(t) \,.
\label{GR1}
\stp
But the KMS condition is that
\bea
G_{F\FF}^<(t) & \equiv & \Tr e^{-\beta H} F\FF e^{iHt} F\FF e^{-iHt}
\nonumber \\ & = &
\Tr e^{-\beta H} \FF e^{-\beta H} e^{\beta H} e^{iHt} F\FF e^{-iHt}
\qquad \mbox{(inserting $e^{\pm \beta H}$)}
\nonumber \\ & = &
\Tr e^{-\beta H}\: e^{\beta H}e^{iHt} F\FF e^{-iHt} e^{-\beta H} F\FF
\qquad \mbox{(cyclicity of the trace)}
\nonumber \\ & = &
\Tr e^{-\beta H}\: e^{i(-i\beta+t) H} F\FF e^{-i(-i\beta+t)H} F\FF
\nonumber \\ & = & G_{F\FF}^>(t-i\beta) \,.
\eea
Substituting into \Eq{GR1} and noting that
$e^{i\omega_{\sss E} \beta} = 1$, we find
\bea
G^{\sss R}(i\omega_{\sss E}) & = &
\int_0^\infty dt e^{i\omega_{\sss E} it} G^>(t)
- \int_{-i\beta}^{\infty -i\beta} dt e^{i\omega_{\sss E} it} G^>(t)
\nonumber \\ & = &
\int_0^{-i\beta} dt e^{i\omega_{\sss E} it} G^>(t)
= -i \int_0^\beta d\tau e^{i\omega_{\sss E} \tau} G^{\sss E}(\tau)
\equiv -i G^{\sss E}(\omega_{\sss E})
\eea
the Fourier (series) transform of the Euclidean function.

The conclusion is that, while Euclidean techniques determine
$G^>(t=0,\x)$ in a straightforward way, the zero Euclidean frequency
correlation function [the Euclidean topological susceptibility]
corresponds to the zero Minkowski frequency,
{\sl retarded} function, not the zero frequency Wightman function.
That is, the Euclidean correlation function continues to a Minkowski
function with the wrong operator ordering.

The Euclidean
function does have {\sl some} relation to the Wightman function, through
the Kramers-Kronig relations.  Namely, defining
\st
\sigma_{F\FF}(t) \equiv G^>(t) - G^<(t) \quad \Rightarrow \quad
\sigma_{F\FF}(\omega) = ( 1- e^{-\beta \omega} ) G^>(\omega)
\simeq \beta \omega G^>(\omega) \,,
\stp
the retarded function at arbitrary complex frequency $\omega$ is
determined by $\sigma(\omega')$ at real frequency by
\st
G^{\sss R}_{F\FF}(\omega)
= \int_{-\infty}^\infty \frac{d\omega'}{2\pi i}
\frac{\sigma_{F\FF}(\omega')}{\omega'-\omega-i\epsilon} \,.
\stp
Evaluating for imaginary (Euclidean) $\omega$ and inverse transforming
to $\tau$,
\st
G^{\sss E}_{F\FF}(\tau)
= \int_0^\infty \frac{d\omega'}{\pi}
\frac{\sigma_{F\FF} \cosh\Big( \omega'(\tau-\beta/2) \Big)}
{\sinh \omega' \beta/2} \,.
\stp
Unfortunately, what we can measure on the lattice,
$G^{\sss E}_{F\FF}(\tau)$, determines an integral over what we want,
$\sigma_{F\FF}(\omega')$.  To proceed further we would need some method
to invert this integral relation.  This problem has frustrated the
evaluation of similar equilibration rates such as the electrical
conductivity and the shear viscosity
\cite{Meyer}.  

We now argue that the problems encountered
in continuing $F\FF$ from lattice data are more severe than for
continuing the correlation functions needed for conductivity or
viscosity.  The problem is that there is no local%
\footnote{By local here we mean that the operator has compact support
  which is independent of the lattice configuration, {\it i.e.}\ it
  stretches for a fixed finite number of lattice spacings.  It may be
  possible to meet our requirements with an operator which falls off
  exponentially with lattice distance, but only if the exponential
  falloff rate is conditional on the smoothness of the lattice
  configuration.}
lattice definition of the topological charge density operator $\FFdual$
which satisfies the property
that $\int d^4 X \FFdual$ over a compact 4-volume is topological (for a
discussion see \cite{MooreMu}).  There {\em are} definitions of
$\int d^4 X \FFdual$ which are topological (given some mild constraints
on the smoothness of lattice fields)
\cite{Luscher}.  That means that there is no problem in
principle with evaluating $G^{\sss E}_{F\FF}(\omega_{\sss E}=0,\p=0)$
or [given real-time lattice configurations] $G^>_{F\FF}(\omega=0,\p=0)$.
But any attempt to compute
$\int d^3 \x \, G^{\sss E}_{F\FF}(\tau,\x)$ relies on measuring $F\FF$
on a 3-D rather than 4-D surface, and so will necessarily be
contaminated by either the non-topological or the non-local
nature of the operators used in the
evaluation.    We believe that this additional complication will make
$\Gamma_{\rm sphal}$ even harder than other transport coefficients to
extract from Euclidean lattice calculations.

\section{Effective theories and computation}
\label{sec:bodeker}

As reviewed in the last section, the sphaleron rate is an intrinsically
Minkowski quantity which cannot be evaluated with Euclidean methods.
Because $\FFdual$ is a total derivative, the sphaleron rate is a
strictly nonperturbative quantity which vanishes to all orders of
perturbation theory. Therefore its evaluation requires
nonperturbative, Minkowski tools.  These are in very short supply.  To
our knowledge, we have only two approaches to measuring the sphaleron
rate:
\begin{itemize}
\item
In certain gauge theories which possess gravity duals (generally highly
supersymmetric theories) there are holographic methods which allow the
sphaleron rate to be related to string theory correlators.  These can be
evaluated on the string theory side in the strong-coupling, large $\nc$
limit of the gauge theory, see \cite{Son_SYM}.  The results are
suggestive but do not apply to ordinary QCD.
\item
In the weak coupling (high temperature) domain, the only nonperturbative
physics which is not exponentially suppressed is essentially {\sl
  classical field} physics.  Therefore it should be possible to evaluate
the sphaleron rate by studying classical field dynamics as an effective
IR description
\cite{Grigoriev,Ambjorn}.
\end{itemize}
In this work we will follow this second approach.  This method has
already been pushed to its logical conclusions for the electroweak
${\rm SU}_{\sss L}$(2)
sector \cite{AmbjornKrasnitz,MooreTurok,HuMuller,ASY1,ASY2,Bodeker,BMR}.
Here we briefly review what that literature has found.

\subsection{B\"odeker's Effective Theory}

Naively, the infrared fields in a weakly coupled quantum field theory of
massless degrees of freedom should obey the classical equations of
motion, in our case
\be
D_\nu F^{\nu\mu} = - J^\mu
\label{Ampere1}
\ee
with the initial field values drawn from a statistical distribution
of classical fields weighted by the Boltzmann factor $\exp(-E/T)$.
B\"odeker has shown \cite{Bodek_hbar} that the first corrections to this
picture enter at order $\hbar^2$ {\sl except} for the possibility of
large (loop) corrections in the ultraviolet.  That is, $J^\mu$ is a
composite operator so an infrared field component could arise from the
overlap of UV fluctuations with almost the same wave number, {\it e.g.\ }
$J^\mu(k) = \int_p \bar\psi(-p) \gamma^\mu \psi(p+k)$.  These fields can
in turn be correlated with past and present IR gauge fields, leading to
effective modifications of the infrared gauge dynamics.  Such large
modifications of the IR dynamics do indeed occur; they are just the
Hard Thermal Loops of Braaten and Pisarski
\cite{Braaten,FrenkelTaylor}.

The sphaleron rate depends in particular on the behavior of transverse
(magnetic) fields in the deep infrared.  In this case the Hard Thermal
Loops have a particularly simple interpretation; the spatial current in
\Eq{Ampere1} is determined by Ohm's law and the color conductivity;
writing noncovariantly, the spatial components read
\cite{ASY2}
\be
D_j F_{ji} + D_t E_i = - J_i \quad = -\sigma E_i + \zeta_i
\label{Langevin1}
\ee
with $\sigma$ the color conductivity and $\zeta_i$ a Langevin noise source
which maintains thermal equilibrium.  In general the color conductivity
is frequency and momentum dependent, but for the most infrared fields of
relevance to the sphaleron rate, ``color randomization'' means that
$\sigma$ should be essentially a constant of value
\cite{Bodeker,ASY2,ArnoldYaffe}
\bea
\label{sigma_value}
\sigma^{-1} & = & \frac{3 \nc g^2 T}{4\pi \mD^2} \left[
   \ln \frac{\mD}{\gamma} + 3.0410 \right] \,,       \nonumber \\
\gamma & = & \frac{\nc g^2 T}{4\pi} 
       \left( \ln \frac{\mD}{\gamma}+\OO(1) \right) \,, \\
\mD^2 & = & \frac{2\nc + \nf}{6} g^2 T^2 \,.
\eea
(Physically, the conductivity $\sigma = \mD^2 / 3\gamma$
is a polarizability times a mean free
path.  The factor of $\mD^2/3$ is the polarizability and $1/\gamma$ is
the mean free path for color randomization.)
This conductivity is parametrically large and allows one to neglect
the $D_t E_i$ term in \Eq{Langevin1} relative to the
$\sigma E_i$ term, yielding the effective description
\be
D_j F_{ji} = - \sigma E_i + \zeta_i \,,\qquad
\int dt \zeta_i(\x,t) \zeta_j(y,0) = 2\sigma T \delta_{ij} 
          \delta^3(\x-\y) \,.
\label{Langevin2}
\ee
The normalization of $\zeta_i$ ensures that the
fields are distributed according to the classical thermal ensemble, as
required by the fluctuation dissipation theorem
\cite{ASY2}.

Note that in finding this effective description and in particular that
$\sigma$ is a constant, we have made an expansion (to second order
\cite{ArnoldYaffe}) in $\ln(1/g)$ the logarithm of the inverse
coupling.  While the numerical coefficient $3.0410$ stabilizes this
expansion somewhat, it is a rather marginal expansion and one should
expect rather large corrections at physically interesting couplings.
Nevertheless we first pursue this approach because the effective theory
involved, \Eq{Langevin2}, is a Langevin equation with very good
mathematical properties.  In particular, unlike \Eq{Ampere1},
\Eq{Langevin2} possesses {\sl dynamics} which have a well defined limit
as the UV regulator is taken to infinity \cite{ASY2,MooreB}.  Therefore
it can be implemented on a lattice, and the lattice spacing
$a\rightarrow 0$ limit may be taken to obtain a well defined limit for
$\Gamma_{\rm sphal}$ in this effective theory.

\subsection{Numerical implementation}

Our implementation of SU(3) gauge theory on a lattice is standard, and
we measure Chern-Simons number using the algorithm of
\cite{brokenphase}, which extends easily to general SU($\nc$).  We do
     {\sl not} follow the implementation of Langevin dynamics from
\cite{MooreB}, however, so we will make some remarks about how we do
implement Langevin dynamics.  First, it is convenient to rescale the
time variable in \Eq{Langevin2} to conventional Langevin time
\be
\label{tau}
\taul = t/\sigma
\ee
which, note, has units of time squared.  In terms of this Langevin time
the fields evolve according to conventional Langevin equations; in
$A_0=0$ gauge,
\be
\label{Langevin3}
\partial_\taul A_i = - D_j F_{ji} + \zeta'_i
= -\frac{\partial H}{\partial A_i} + \zeta'_i \,,
 \qquad
\int d\taul \zeta'_i(\taul,y) \zeta'_j(x,0) = 
    2T \delta_{ij} \delta^3(\x-\y)\,.
\ee
The sphaleron rate $\Gamma_{\sss L}=\int d\taul \int d^3 \x
F\FF(\taul,\x) F\FF(0,0)$ obtained in this theory is related to that
obtained using \Eq{Langevin2} by a factor of $\sigma$.

Any numerical implementation of \Eq{Langevin3} requires discretization
of $\taul$ into finite steps.  The challenge is that larger steps result
in a faster algorithm but may have larger numerical errors; in
particular the sphaleron rate scales as $T^4$ and so is sensitive to any
correction which shifts the thermodynamics.  Therefore we particularly
want an algorithm which respects the correct thermodynamics to good,
controllable accuracy.  Our choice is to implement \Eq{Langevin3} as a
series of short Hamiltonian trajectories of length $\tham$,
with the electrical field
drawn randomly from a Gaussian ensemble at the beginning of each
trajectory.  To see how this reproduces \Eq{Langevin3}, consider the
following Hamilton's equations (in $A_0=0$ gauge, which we choose for
explanatory convenience):
\bea
\partial_t E_i & = & -\frac{\partial H}{\partial A_i} \,,
\nonumber \\
\partial_t A_i & = & E_i \, ,
\nonumber \\
\langle E_i(\x,t=0) E_j(\y,t=0) \rangle & = & T \delta_{ij}
 \delta^3(\x-\y) \,.
\eea
(The lattice implementation of these equations is known, see for
instance \cite{Ambjorn} but note that in the above we do {\em not}
enforce Gauss' Law.)
To second order in small $\tham$, the change in $A_i$ is found to be
\be
A_i(\x,\tham) - A_i(\x,0) 
= E_i(\x) \tham - \frac{\partial H}{\partial A_i} \frac{\thamsq}{2}
    + \OO(\tham^3)\,.
\ee
This is the same as \Eq{Langevin3} if we interpret $\taul = \thamsq/2$.
Hence, a randomization of the $E$ fields followed by an evolution of
time $\tham$ corresponds to an evolution of $\taul=\thamsq/2$ units of Langevin
dynamics up to higher order in $\tham$ corrections.  Higher order
corrections are inevitable in a discretization of Langevin dynamics, and
in this case we know that any corrections to the interpretation of the
algorithm in terms of Langevin dynamics must enter at even powers in
$\tham$, since the algorithm described is actually $t$-even.  In practice we
measure $\Gamma_{\sss L}$ for two or more
values of $\tham$ and make an extrapolation to the small $\tham$ limit,
based on values no longer than one lattice unit.
We also apply the known $\OO(a)$ lattice-continuum corrections to the
lattice spacing \cite{orderA2} and to the time scale $t$, which is known
for Hamiltonian dynamics \cite{MooreB}.

\section{Weak-coupling Results}
\label{results}

We want the large volume, small lattice-spacing limit and our goal is a
few percent precision.  There are many systematic limits to consider;
lattice spacing $a\rightarrow 0$, volume $V\rightarrow \infty$,
refresh frequency $\tham \rightarrow 0$, leapfrog timestep size, and
parameters of
our $\ncs$ measuring method.  Some of these are simple to control; we
can check systematically that our $\ncs$ measurement is topological
because our method is ``calibrated'' with integrations to the vacuum
\cite{brokenphase}.  Errors associated with the leapfrog step size are
quadratic in the step; we compared step sizes of 0.2 and 0.1 and found
a $4\pm 2\%$ change in $\Gamma_{\sss L}$.  Hence the difference between
0.1 and 0 is
$1.3\pm 0.7\%$ which is acceptably small.  We typically extrapolate over
two $\tham$ values, $\tham=a$ and $\tham=a/2$.  The resulting
$\Gamma_{\sss L}$ generally differ by less than
$10\%$, so higher orders
in the extrapolation should be at the $1\%$ level and therefore
negligible.

\begin{figure}
\centerbox{0.75}{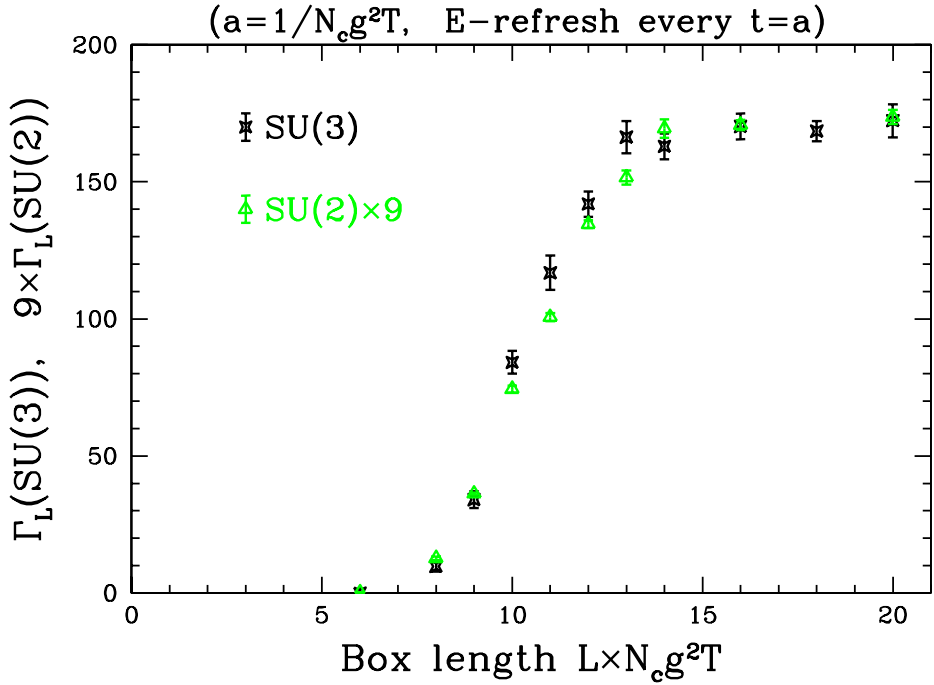}
\caption[finite volume scaling]
{\label{fig_vol} (Color online)
Volume dependence of the sphaleron rate in Langevin dynamics,
determined on lattices with spacing $a = 1 / \nc g^2 T$, for
SU(2) and SU(3).  Errors are statistical only; data are not extrapolated
to $t=0$ or $a=0$.  Using the indicated scaling,
the SU(2), SU(3) behaviors look very similar; the rate
saturates around $L=15/\nc g^2 T$.}
\end{figure}

Next we consider the extrapolation to infinite volume.  The good news is
that 3-D SU($\nc$) gauge theory has a mass gap, so the approach to large
volume behavior is expected to improve exponentially in box length for
boxes much larger than the correlation length.  In practice we measure
the sphaleron rate at fixed lattice spacing $a$
(technically, at a fixed value of the dimensionless ratio $g^2 a T$) for
several volumes to see what volume is sufficient to achieve the large
volume limit.  This was first done by Ambj{\o}rn and Krasnitz
\cite{AmbjornKrasnitz}.  Our results, shown in Fig.~\ref{fig_vol},
indicate that the volume dependence of the sphaleron rate in SU(2) and
SU(3) gauge theory are almost identical {\em if} we scale the box length
by $\nc g^2 T$, not $g^2 T$.  We also see that the sphaleron rate
increases with volume but saturates around
$L = 15 / \nc g^2 T$.  Therefore we should make sure
$L > 15 / \nc g^2 T$  to be in the infinite volume limit.  We therefore
use $L \geq 18 / \nc g^2 T$ in what follows.%
\footnote{Note that using very large volumes is not helpful;
  since the measurement of $\ncs$ is global over the lattice, there is
  no gain in statistics as the lattice volume increases, while the
  numerical effort increases with volume.}

Finally, we consider the small lattice spacing $a$ limit.  Since the
numerical effort required to get the same level of statistical
significance increases as $a^{-5}$, we should try to understand this
limit as well as possible.  We know that, at the level of the
thermodynamics of the system, there are corrections already at the
$\OO(a)$ level.  These are well understood and amount to an effective
rescaling of the lattice spacing \cite{orderA2}:
\begin{equation}
\frac{1}{a_{\rm corrected}} = \frac{1}{a_{\rm bare}}
- \nc g^2 T \times \left\{ \begin{array}{ll}
0.07918 & \mbox{SU(2)} \\
0.10232 & \mbox{SU(3)} \\
0.12084-1/(6\nc^2) & \mbox{SU($\nc$)} \\
\end{array} \right.
\end{equation}
Choosing $a < 1 / \nc g^2 T$ is then sufficient to push this correction
under $10\%$.  There are unknown $\OO(a^2)$ corrections which are
presumably of order the square of this size, so $a < 1/\nc g^2 T$ would
leave of order $1\%$ unknown corrections to the effective lattice
spacing.  But the scaling between lattice and continuum Langevin
diffusion rates involves $a^5$, so even a $1\%$ error in $a$ leads to a $5\%$
error in the diffusion rate, which is not acceptable.
There are also other unknown $\OO(a^2)$ corrections which
cannot necessarily be represented as a rescaling of $a$ or of $\taul$.
Hence we should try to get to lattices which are somewhat
smaller than $a = 1/\nc g^2 T$, and to perform an $a\rightarrow 0$
extrapolation over the remaining $a^2$ corrections.

\begin{figure}
\phantom{.} \hfill \putbox{0.46}{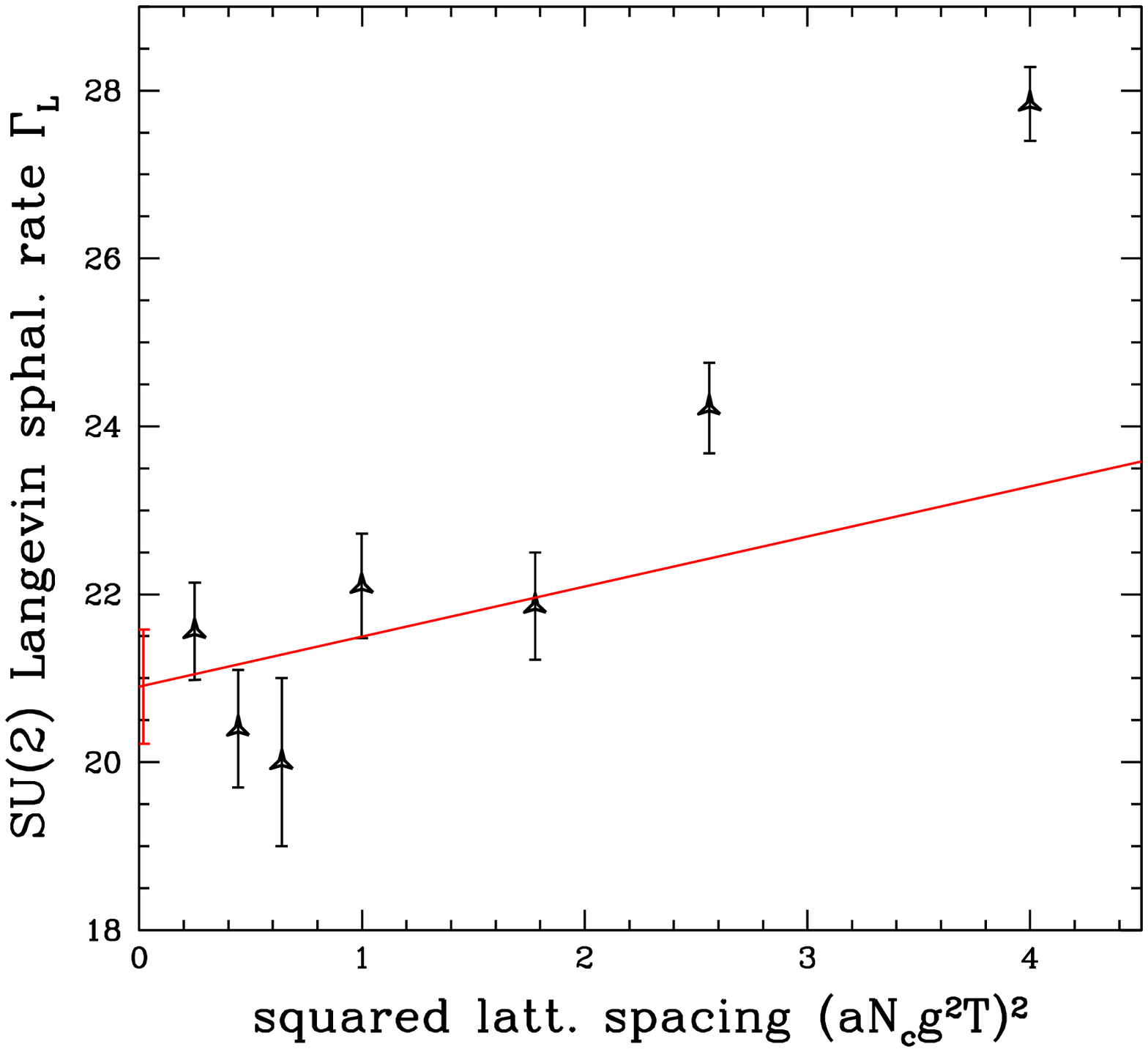} \hfill
\putbox{0.46}{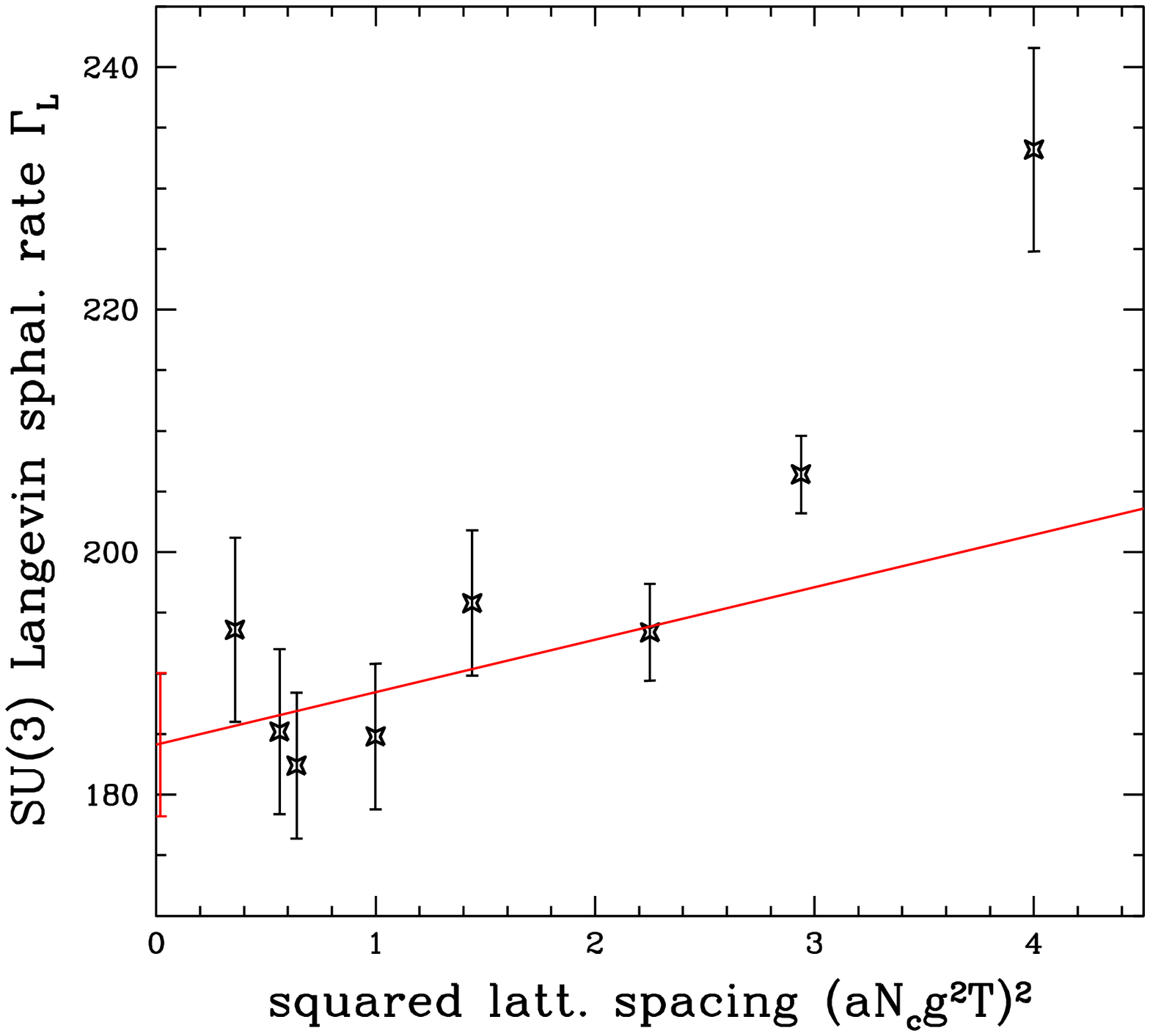} \hfill \phantom{.}
\caption[Finite lattice spacing behavior]
{(color online) Finite lattice spacing extrapolation of SU(2) [left] and
  SU(3) [right] lattice Langevin
  sphaleron rate.  The two largest lattice-spacing points were not
  used in the extrapolation.  \label{fig:result1}}
\end{figure}

\def\ssp{\hspace{2.7em}}

\begin{table}
\centerline{\begin{tabular}{|ccc|ccc|} \hline
$\;\;$Group$\;\;$ & $\; a \nc g^2 T \;$ & $\ssp \Gamma_{\sss L} \ssp$ &
$\;\;$Group$\;\;$ & $\; a \nc g^2 T \;$ & $\ssp \Gamma_{\sss L} \ssp$ \\ \hline
SU(3) & 2.00 & $233.2\pm 8.4$ & SU(2) & 2.00 & $ 27.84\pm 0.44$ \\
SU(3) & 1.72 & $206.4\pm 3.2$ & SU(2) & 1.60 & $ 24.22\pm 0.54$ \\
SU(3) & 1.50 & $193.4\pm 4.0$ & SU(2) & 1.33 & $ 21.86\pm 0.64$ \\
SU(3) & 1.20 & $195.8\pm 6.0$ & SU(2) & 1.00 & $ 22.10\pm 0.62$ \\
SU(3) & 1.00 & $184.8\pm 6.0$ & SU(2) & 0.80 & $ 20.00\pm 1.00$ \\
SU(3) & 0.80 & $182.4\pm 6.0$ & SU(2) & 0.67 & $ 20.40\pm 0.70$ \\
SU(3) & 0.75 & $185.2\pm 6.8$ & SU(2) & 0.50 & $ 21.56\pm 0.58$ \\
SU(3) & 0.60 & $193.6\pm 7.6$ &       &      &
\\ \hline
\end{tabular}}
\caption[Langevin rate as function of $a$]
{Numerical results for the sphaleron rate under Langevin diffusion,
for SU(2) and SU(3) gauge theory and several lattice spacings.  Each
result is taken at a volume with $L \geq 18 / \nc g^2 T$ and is already
extrapolated to zero $\tham$; systematic errors from these
extrapolations are smaller than the indicated statistical errors.
\label{table:result1}}
\end{table}

Our data for $\Gamma_{\sss L}$ as a function of $a^2$ is presented in
Fig.~\ref{fig:result1} and Table \ref{table:result1}.  The two
largest-$a$ lattices used are not included in the fit, because the
corrections are too large (the fit becomes poor) and because the
behavior of $\ncs$ is not yet
strictly topological (the $\ncs$ measuring algorithm sometimes shows an
inconsistency in which $\int F\FF$ from vacuum, along the Langevin
history, and back to vacuum is far from an integer).  To convert the
extrapolated value into the result in \Eq{mainresult}, we multiply
by $(3\nc/4\pi)$ as indicated in \Eq{sigma_value}, \Eq{tau}.
As a check on our Langevin dynamics method, we have also repeated the
calculation of the SU(2) sphaleron rate originally performed in
\cite{MooreB}; we find that the coefficient in \Eq{mainresult}
for SU(2) is $10.0 \pm 0.3$, in agreement with $10.8\pm 0.7$ found
there.

\begin{table}
\centerline{\begin{tabular}{|c|c|c|} \hline
$\;$ Colors $\nc\;$ & $\;$ Coeff.\ in \Eq{mainresult} $\;$ &
$\;$ Coeff.\ in \Eq{nc_scale} $\;$ \\
\hline
2 & $9.53 \pm 0.14$  & $0.198  \pm .003$ \\
3 & $127.8 \pm 4.0$  & $0.197  \pm .006$ \\
5 & $2790 \pm 70$    & $0.186  \pm .005$ \\
8 & $51300 \pm 3100$ & $0.199  \pm .012$ \\ \hline
\end{tabular}}
\caption[Langevin rate as function of $\nc$]
{The Langevin equation $\ncs$ diffusion rate as a function of the number
  of colors $\nc$, for a fixed lattice spacing
$a = 0.8 / \nc g^2 T$ and $E$-field refresh rate $\tham=a$.
\label{table:nc}}
\end{table}

In \cite{MooreSEWM2K} one of us speculated that
$\Gamma_{\sss L}$ should scale with $\nc$ as
$\frac{\nc^2-1}{\nc^2} (\nc \alpha)^5 T^4$, and in particular
\st
\label{nc_scale}
\Gamma_{\rm sphal} = \kappa' \left( \frac{\nc g^2 T^2}{\mD^2} \right)
\left( \ln \frac{\mD}{\gamma} + 3.0410 \right)
(\nc^2-1) \nc^3 \alphas^5 T^4 \,,
\stp
with $\kappa' \sim 0.24$.  We have investigated this speculation
by computing the sphaleron rate at a single value of
$a = 0.8 /\nc g^2 T$ for several values of $\nc$, shown in
Table \ref{table:nc}.  The data in the table are also not extrapolated
to $\tham=0$ but are computed at $\tham=a$.  Extrapolating by assuming that the
$a$ and $\tham$ scaling are the same as for $\nc=2,3$, the $a^2$ correction
lowers the rate by $1.8\%$ and the $\tham\rightarrow 0$ correction
raises the rate $9\%$ so we find $\kappa' = 0.21\pm .01$, as
quoted in \Eq{ncscaling}.

\section{Finite-coupling desperation}
\label{finite_a}

Our calculation has made a strict weak-coupling expansion, which is
really two approximations.  First, in writing the theory in terms of
classical fields on the lattice, we assume the nonperturbative IR
dynamics is essentially classical.  Second, by implementing Langevin
dynamics, we assume that color conductivity is
large and wavelength independent.  We cannot relax the first
approximation without losing our computational method altogether.  But
we can relax the second approximation and allow the conductivity to be
finite, by studying the theory with Hamiltonian dynamics
\cite{Ambjorn,AmbjornKrasnitz}.  This allows us access to
``intermediate'' couplings, though with larger theoretical errors.

At weak coupling, the continuum theory has three parametric length
scales:
\begin{itemize}
\item
The length scale $1/T$, where most thermal degrees of freedom reside,
\item
The length scale $1/\sqrt{\nc} \,gT$ associated with screening phenomena,
\item
The length scale $1/\nc g^2 T$, where physics is nonperturbative.
\end{itemize}
The corresponding scales for the Hamiltonian lattice theory are
$a$, $\sqrt{a/\nc g^2 T}$, and $1/\nc g^2 T$.  Only two of these scales
are relevant for the dynamics of topology change; the $1/\nc g^2 T$ scale
where the nonperturbative physics occurs, and the $1/\sqrt{\nc}\, gT$
``screening'' scale which influences the dynamics of the IR scale.
On the lattice one does not choose $a$, $g^2$, and $T$ separately; one
controls the dimensionless ratio $a \nc g^2 T$
(often written $2\nc^2/\beta_{\sss L}$).  But this choice can be
paraphrased as a choice for a finite value of $g^2 \nc$, as we now show.

Since the sphaleron rate involves the evolution of infrared magnetic
fields, the relevant feature of $1/gT$ physics is the ``damping
rate'' for the evolution of magnetic fields \cite{Arnoldlatt}.
In the continuum this is proportional to $\mD^2$.  On the lattice it is
proportional to an integral which diverges as $1/a$.%
\footnote{In studying the SU(2) sphaleron rate, where the gauge coupling
  is small, it is convenient to add ``extra'' degrees of freedom to the
  lattice, as done in \cite{HuMuller,BMR}.  This effectively shortens
  the screening length scale on the lattice, which is the same as
  effectively making the coupling smaller.  Since we want to explore the
  largest possible couplings, this is the opposite of what we want,
  which is why we consider the pure classical lattice theory without
  additional degrees of freedom.}
Equating them, we find \cite{Arnoldlatt}
\begin{equation}
k \gamma_{\rm contin} = \frac{(2 \nc + \nf) \pi g^2 T^2}{24} =
k \gamma_{\rm latt} = \frac{2.14988 \pi \nc g^2 T}{8\pi a} \,,
\end{equation}
which relates the screening scales of the lattice and continuum
theories.  Re-organizing, we can determine $g^2$ in terms of
$(a \nc g^2 T)$:
\be
\label{match_md}
\frac{2.14988 \times 3}{\pi \frac{2\nc+\nf}{\nc}} \nc g^2
= a \nc g^2 T  \quad \Rightarrow \quad
\frac{g^2}{4\pi} = \frac{2\nc + \nf}{12 \nc^2 (2.14988)} (a \nc g^2 T)
\,.
\ee
We use this relation to convert a value of the lattice spacing --
really, of $a \nc g^2 T$ -- into an equivalent value of
$\alphas = g^2/4\pi$.

\begin{table}
\centerline{\begin{tabular}{|c|c|c|c|c|} \hline
$\; a \nc g^2 T \;$ & $\; \alphas$(0 flavor) $\;$ &
$\;\alphas$(3 flavor)$\;\;$ &
$\;\;\Gamma_{\rm sphal} / \alphas^4 T^4 \;\;$ &
$\Gamma_{\rm sphal}/\alphas^4 T^4$ from Eq.(\ref{mainresult})  \\ \hline
2.40 &  0.062  &  0.093 &  $17.28 \pm 0.30$   & 27.7 \\
2.00 &  0.052  &  0.078 &  $14.06 \pm 0.20$   & 23.5 \\
1.72 &  0.044  &  0.066 &  $12.60 \pm 0.27$   & 20.5 \\
1.50 &  0.039  &  0.058 &  $11.45 \pm 0.26$   & 18.2 \\
1.20 &  0.031  &  0.047 &  $10.41 \pm 0.22$   & 14.9 \\
1.00 &  0.026  &  0.039 &  $9.18  \pm 0.23$   & 12.7 \\
0.75 &  0.019  &  0.029 &  $8.51  \pm 0.16$   &  9.8 \\
0.60 &  0.016  &  0.023 &  $7.81  \pm 0.20$   &  8.0 \\ \hline
\end{tabular}}
\caption[Hamiltonian Sphaleron rate]
{\label{table_finitea} SU(3) sphaleron rate under Hamiltonian dynamics,
  and its interpretation in terms of $\alphas$ for the 0-flavor and
  3-flavor theories.  Errors shown are statistical, and are dwarfed by
  much larger theoretical errors in equating lattice and continuum
  theories, which we have not estimated.  We also show the result
  implied by \Eq{mainresult}, which works well at the smallest lattice
  spacings but is a poor description for the coarsest lattices.}
\end{table}

Table \ref{table_finitea} presents our results for the sphaleron rate
under Hamiltonian dynamics.  The largest value of $a \nc g^2 T$
corresponds to the point where we can barely distinguish topological
behavior on the lattice.  For this value, the infinite volume limit was
achieved with a lattice 8 on a side!  For $\alphas$ larger than this
corresponding value, our ability to guess the strong sphaleron rate gets
even weaker; we must extrapolate based on the behavior shown in the
table.  This corresponds to the case relevant for heavy ion collisions.
The figure also shows the performance of the leading-log expansion, that
is, the guess for the sphaleron rate using \Eq{mainresult} together with
the matching for the Debye screening scale from \Eq{match_md}.  The
leading-log (B\"odeker effective theory) approach reproduces the lattice
Hamiltonian result for the finest lattices but is rather far off for the
coarsest lattices, showing that the evolution is not overdamped for such
coarse lattice spacing (large $\alphas$).

\section{Discussion}
\label{discussion}

We have computed the sphaleron rate in SU(3) gauge theory at weak
coupling.  In the weak coupling limit, there is a rigorous relation
between the sphaleron rate and the topological diffusion rate for
classical lattice gauge theory under Langevin dynamics, which we have
computed with few-\% error bars.  We have also shown that the
$\nc$-dependence for the sphaleron rate has a surprisingly simple and
intuitive behavior.  Our analysis of Hamiltonian dynamics
on coarser lattices suggests that this B\"odeker effective description
only works at very small values of $\alphas$.  While it should apply for
the SU(2) electroweak sector and for QCD at GUT scale temperatures, it
is not a successful description at electroweak temperatures.

For applications to electroweak baryogenesis, we need the
strong sphaleron rate in an $\nf=6$ theory with $\alphas \sim 0.1$.  The
value of $\alphas$ for $\nf=6$ is twice the value for $\nf=0$, so this
corresponds roughly to our $(a\nc g^2 T)=2$ data point
in Table \ref{table_finitea}; the strong
sphaleron rate relevant at the electroweak scale is approximately
\st
\Gamma_{\rm strong \: sphal}(\nf=6,\alphas\sim 0.1)
\simeq 14 \alphas^4 T^4 \,.
\stp
It is difficult to assign theoretical error bars to this estimate but
they should be rather large; the relevant lattice spacing is so large
that the topological nature of the $F\FF$ measurement is barely under
control, and the lattice spacing corrections discussed earlier are not
small.  It is also not clear that the lattice field dynamics are a good
description of the continuum, not-so-small coupling dynamics.  We only
have rigorous arguments that they agree for small coupling.%
\footnote{Actually even the subleading in log correction 3.041 of
\Eq{mainresult} will be different on the lattice.  The lattice value for
this constant has not been computed.}  The
continuum dynamics have $\OO(g)$ (technically $\OO(\nc g^2 T/\mD)$)
corrections, the lattice dynamics have $\OO(\sqrt{\nc g^2 aT})$
corrections, and there is no guarantee that they are the same
(see for instance \cite{LMPT}).  Therefore we expect at least
$\OO(50\%)$ systematic corrections to this result.

Unfortunately, the coupling values $\alphas \sim 1/3$ relevant for heavy
ion physics correspond to lattices too coarse for any study of topology
to be successful.  Therefore the best we can do is to extrapolate our
existing results towards this regime.  One might guess based on
Table \ref{table_finitea} that the sphaleron rate at large couplings
is $\Gamma_{\rm sphal} \sim 30 \alphas^4 T^4$.  But this is a desperate
extrapolation.  It is not clear to us how to proceed at such large
couplings; as we have argued, analytic continuation from Euclidean
lattice calculations seems to be even more difficult than for other
transport coefficients.  But at least the behavior at the electroweak
temperature scale, and at higher temperatures, is under some control.

\section*{Acknowledgements}

We would like to thank Dima Kharzeev, for stimulating us to revisit this
issue after so many years.
This work was supported in part by the Natural Sciences and Engineering
Research Council of Canada.

\end{document}